\headline={\ifnum\pageno>1 \hss \number\pageno\ \hss \else\hfill \fi}
\pageno=1
\nopagenumbers
\vskip 55mm
\centerline{\bf AN EXPLICIT CONSTRUCTION OF CASIMIR OPERATORS}
\centerline{\bf AND EIGENVALUES : I }
\vskip 25mm
\centerline{\bf H. R. Karadayi and M. Gungormez}
\centerline{Dept.Physics, Fac. Science, Tech.Univ.Istanbul }
\centerline{ 80626, Maslak, Istanbul, Turkey }
\vskip 10mm
\centerline{\bf{Abstract}}
\vskip 10mm
We give a general method to construct a complete set of {\bf linearly
independent} Casimir operators of a Lie algebra with rank N. For a Casimir
operator of degree \ p, this will be provided by an explicit calculation of
its symmetric coefficients $ g^{A_1 A_2.. A_p} $. It is seen that these
coefficients can be described by some rational polinomials of rank ~N. These
polinomials are also multilinear in Cartan sub-algebra indices taking values
from the set $ I_\circ \equiv \{1,2, .. , N\} $. The crucial point here is that
for each degree {\bf one needs, in general, more than one polinomial}. This
in fact is related with an observation that the whole set of symmetric
coefficients $ g^{A_1 A_2.. A_p} $ is decomposed into some sub-sets which
are in one-to-one correspondence with these polinomials. We call these
sub-sets {\bf clusters} and introduce some {\bf indicators} with which
we specify different clusters. These indicators determine all the clusters
whatever the numerical values of coefficients $ g^{A_1 A_2.. A_p} $ are.
For any degree p, the number of clusters is independent of rank N. This
hence allows us to generalize our results to any value of rank N.

To specify the general framework, explicit contructions of 4th and 5th order
Casimir operators of $A_N$ Lie algebras are studied and all the polinomials
which specify the numerical value of their coefficients are given explicitly.

\vskip 40mm
\vskip 40mm

\footnote{}{e-mail: karadayi@sariyer.cc.itu.edu.tr}

\hfill\eject

\vskip 3mm
\par {\bf{I. INTRODUCTION}}
\vskip 3mm
For a classical or exceptional Lie algebra $\bf G$, the problem of finding
explicit expressions for Casimir operators and their eigenvalues is of
principal importance both in physics and in mathematics. A Casimir operator
$\bf I(p)$ of degree p can be expressed by
$$ I(p) \equiv \sum_{A_1 \leq A_2 \leq ... \leq A_p} g^{A_1,A_2.. A_p} \
sym( T_{A_1} T_{A_2} .. T_{A_p} ) \eqno(I.1) $$
where the sum is over indices taking values from the set
$$ S \equiv \{1,2, .. dim\bf G \} \ \ .  \eqno(I.2)  $$
The coefficients $ g^{A_1,A_2.. A_p} $ can be assumed to be completely
symmetric and hence  $\bf sym(...)$ means complete symmetrization with
weight $\bf{1}$ for the generators $ T_A $ of \ {\bf G}. For a D-dimensional
representation, corresponding eigenvalues can then be expressed by
$$ {1\over{\bf{D}}} \ \  Trace({\bf I(p)}) \ \ . \eqno(I.3) $$

The expression (I.1) is in principle due to Poincare-Birkhoff-Witt {\bf(PBW)}
theorem while (I.3) is a result of Schur lemma {\bf [1]}. One must note
here that (I.1) has an abstract meaning which is valid for any representation
of {\bf G} whereas (I.3) must be calculated for each representation separately.
In physics litterature, there are several works {\bf [2]} dealing
with appropriate descriptions of these two expressions. We also emphasize the
works of Okubo and Patera {\bf [3]} concerning a study of fourth and
fifth order Casimirs which we now consider as preliminaries to our work.
Extensions to superalgebras are made in some relatively recent works {\bf [4]}
and there are recent efforts for generalizations to q-superalgebras {\bf [5]}.
For affine Kac-Moody algebras, an extension of the generalized second order
Casimir element to higher orders has also been accomplished {\bf [6]}.

\par On the other hand, the results of all these efforts find applications
in string theories. A spectacular example is the well-known anomaly
cancellation mechanism {\bf [7]} which is based on some very special non-linear
relationships among eigenvalues (I.3) of operators (I.1). These were for
the days of first string revolution. In the present days of second string
revolution {\bf [8]}, it is seen that all Casimir invariants are involved
in a highly non-trivial way . For instance, the relevant operators of N=2
supersymmetric models {\bf [9]} can be expressed in terms of Casimir operators.

\par Beside applications, it is known {\bf [2]} that a complete
knowledge for a Lie algebra can be obtained with a complete knowledge for all
its independent Casimir operators and their eigenvalues. Hence, for all
irreducible representations, an actual calculation of (I.3) is a worthwhile
task in principle as well as in practical applications. To this end, one of
the essential problems is to calculate the multiplicity of weights
participating in representations and this problem is also at the root of Weyl
character formulas or recursive multiplicity formulas which are due to
Freudenthal and Kostant. Explicit calculation of traces could still be
complicated for irreducible matrix representations having higher dimensions.
This will be considered in the second part of our work which is based on
the mechanism presented in a previous article {\bf [10]}.

\par Another problem appears in the determination of the relationships which
are known to exist among different Casimir operators of the same Lie algebra.
This is indeed a natural problem when one considers that Casimir elements are
generally defined on the universal enveloping algebra with a \ PBW basis.
As is known {\bf [1]}, a \ PBW  basis is formed by monomials constructed from
multiple products of Lie algebra generators. It is therefore natural to ask
some relationships among different Casimir elements if one wants only a finite
number of them.  One can think such relationships in two ways:
\vskip 2mm
\par (i) Non-linear dependences among Casimir elements of different orders,
\vskip 1mm
\par(ii) Linear dependences among Casimir elements of the same order.
\vskip 2mm
\noindent Due to illuminating works of Borel and Chevalley {\bf [11]} and
possibly some others working on group cohomologies, the problem (i) has been
solved with the calculation of {\bf Betti numbers} of topological spaces formed
by group manifolds. These Betti numbers are known, in the same time,
to be {\bf exponents} which specify the degrees of Racah invariants {\bf [12]}.
The Racah invariants also provide us a way to determine the Casimir operators
which are independent non-linearly. Gruber and O'Raifertaigh {\bf [13]} show
us a generalization of these invariants. It is however well-known that all
these constructions are not unique,~i.e. there could always be some other
choices and clearly this has to do with (ii). If one recalls, on the contrary
of our choice in (I.1), that Casimir elements are in general non-homogeneous
superpositions of PBW monomials, it is seen that this last problem has also
some basic features. In two successive works, we will give a unified picture
for these problems. For ${\bf A_N}$ Lie algebras, a general solution to (i)
is as in the following
\vskip 3mm
Let $ {\bf \kappa(p) } $ be the number of linearly independent Casimir
operators expressed in the {\bf homogeneous} form of (I.1). Then,
$$  \kappa(p)= the \ number \ of \ partitions \ of \ p \ into \
all \ positive \ integers \ except \ 1 \ \ . $$
\vskip 3mm
\noindent Some examples will be instructive here:
\vskip 3mm
\par   (i) \ \ \ $ \kappa(4)$ = 2 \ \ due to \ \ 4 = 2 + 2
\vskip 1mm
\par  (ii) \ \ $ \kappa(5)$ = 2 \ \ due to \ \ 5 = 3 + 2
\vskip 1mm
\par (iii) \ $ \kappa(6)$ = 4 \ \ due to \ \ 6 = 4 + 2 = 3 + 3 = 2 + 2 + 2
\vskip 1mm
\par (iv) \ $ \kappa(7)$ = 4 \ \ due to \ \ 7 = 5 + 2 = 4 + 3 = 3 + 2 + 2
\vskip 3mm

Our solutions for the numerical values of coefficients $ g^{A_1,A_2.. A_p} $
have the generic form for p=4,5
$$  P \ parameter(1) + Q \ parameter(2) \ \ . \eqno(I.4)  $$
As will be shown in the following sections, the choice of two free parameters
in (I.4)  will be made as being in line with the partitions 4 = 2+2 or 5 = 3+2.
The coefficients P and Q are {\bf rational} polinomials of rank \ N \ and
they are also multi-linear in Cartan sub-algebra indices  $i_1,i_2, .. i_p$
taking values from the set ${\bf I_\circ} \equiv \{ 1,2, .. N \} $. It is
crucial to note here that {\bf we always need several P and Q polinomials
in order to describe the whole set of coefficients ${\bf g^{A_1,A_2.. A_p} }$}.
To this end, we introduce two novel concepts: {\bf clusters} and
{\bf indicators}. The clusters are defined to be subsets of coefficients
$ g^{A_1,A_2.. A_p} $ with the same numerical value and the whole set of
these coefficients has a direct sum decomposition in terms of these clusters.
It is seen that a unique polinomial P or Q can be assigned only to a cluster
and hence we need in general more than one polinomial for the whole set of
coefficients. The indicators, on the other hand, are defined in such a way
that they take different values on different clusters and hence they count
the number of different clusters. The procedure which is introduced by the
clusters and indicators works in an independent way from the rank N and this
allows us to extend our results for all values of the rank N.

Within the scope of this work, we expose only the results for p=4,5.
The generalizations beyond p=7 begin to be difficult because the number of
indicators increases and at present we can not be able to find a systematic
procedure to determine all the indicators completely.

In section II, we will give some useful notation and introduce some generalized
scalar products with which we define indicators. Our results for degrees p=4,5
will be given respectively in sections (III) and (IV). Some features will also
be emphasized in the last section.

\vskip 3mm
\par {\bf{II. CLUSTERS AND INDICATORS}}
\vskip 3mm
Let us begin with a detailed description of the generator basis $T_A$ for a
Lie algebra {\bf G} $( \equiv {\bf A_N} )$ which is defined by
$$  [T_A,T_B]= F_{A B}^C ~T_C \ \ . \eqno(II.1) $$
$F_{A B}^C$'s here are structure constants and indices \ A,B,C \ take values
from the set \ S, as in (I.2). We assume it has a triangular decomposition
$ S \equiv S_+ \oplus S_- \oplus S_\circ $ in such a way that:
\vskip 2mm
\par (i)   $S_+ \equiv \{1,2, .. {1 \over 2} N (N+1) \} $
\vskip 1mm
\par (ii)  $S_- \equiv {1 \over 2} N (N+1) \oplus S_+ $
\vskip 1mm
\par (iii) $S_\circ \equiv N (N+1) \oplus I_\circ $

For $i \in I_\circ $, the generators which correspond to simple roots
$\alpha_i$ can be chosen by
$$
e_{\alpha_i} \equiv T_i \ \ , \ \
f_{\alpha_i} \equiv T_{i+{1 \over 2} N (N+1) } \ \ , \ \
h_{\alpha_i} \equiv [e_{\alpha_i},f_{\alpha_i}] = T_{i+N (N+1)} \ \ . \eqno(II.2) $$
and explicit matrix representations are always exist so that
the set $\{ e_{\alpha_i}~,~f_{\alpha_i}~,~h_{\alpha_i} \}$ forms a
{\bf Chevalley basis}. For the whole set of generators, one has a similar
triangular decomposition $ G \equiv G_+ \oplus G_- \oplus G_\circ $ where
$G_\circ $ is a Cartan sub-algebra and $ G_+ \oplus G_\circ $ one of its
Borel sub-algebras. For this and other relevant techniques of Lie algebras,
we will refer the excellent book of Humphreys {\bf [14]}. For any irreducible
representation specified by a dominant weight $\Lambda$, trace operations
can be fixed by
$$ Trace(T_A T_B) = c_2 (\Lambda) ~g_{AB} \eqno(II.3) $$
where $g_{AB}$ is Killing-Cartan metric having the explicit matrix form
$ g = diagonal( X~,~K ) $. Here \ X \ is a $N(N+1) \times N(N+1)$ \ \
dimensional symmetric sub-matrix with non-zero elements
$ X_{a,a+N(N+1)/2} $ = 1 for $a \in S_+ \oplus S_- $ and \ K \ is just the
Cartan matrix. $c_2 (\Lambda)$ here is sometimes called the {\bf second index}
and it is normalized by
$$ c_2 (\lambda_i) = Binomial(N-1,i-1) \eqno(II.4) $$
for elementary representations of $A_N$ Lie algebras. Elementary
representations are defined to be the ones characterized by {\bf fundamental
dominant weights} \ $ \lambda_i $ \ which are duals of simple roots $\alpha_i$.
They consist of only one Weyl orbit and hence exhibit no complications due to
multiplicity problems encountered in their explicit matrix constructions.
The numerical values of structure constants $F_{AB}^C$ are also fixed by
$$ F_{AC}^D F_{BD}^C \equiv 2 ~(N + 1) ~g_{AB} \ \ . \eqno(II.5) $$
Now and then, summation is adopted over the repeated indices. All these
completely determine the normalizations for which we use in the explicit
construction of matrix representations.

The starting point now is
$$ F_{AB}^{\{C_1} ~g^{C_2...C_{p-1} \}B} = 0 \eqno(II.6) $$
which is the result of invariance property of Casimir operators.
Two-indexed solutions of (II.6) is just the inverse
$g^{-1}_{AB} \equiv g^{AB}$ of Killing-Cartan metric. The following properties
of coefficients $ g^{A_1,A_2,..A_p} $ would be quite helpful before solving
(II.6) explicitly for $p \ge 4$ :
\vskip 3mm
$ \underline {Invariance} $ : $ g^{A_1,A_2,..A_p} $'s \ are non-zero only for
$$ \alpha_{A_1} + \alpha_{A_2} + ... + \alpha_{A_p} = 0 $$

where \ $\alpha_I \equiv 0 $  for  $ I \in S_\circ $ ,
\vskip 3mm
$ \underline {Parity} $ : $ g^{A_1,A_2,..A_p} \equiv g^{P(A_p),P(A_{p-1}),..P(A_1)} $
\ \ where \ \ $\alpha_{P(A)} \equiv -\alpha_A$ ,
\vskip 3mm
$ \underline {Duality} $ : $ g^{A_1,A_2,..A_p} \equiv (-1)^{\sigma}
g^{{A_1}^\star,{A_2}^\star,..{A_p}^\star} $ \ where $\alpha_{A^\star}$
is just the conjugate of $ \alpha_A $ under diagram automorphism of $A_N$
Lie algebras and  $\sigma$ is a real phase which will be specified later ,
\vskip 3mm
$ \underline {Weyl \ Symmetry} $ : Let us consider the action of Weyl group
{\bf W}: $\Theta(\alpha_a) \equiv \alpha_{\theta(a)}$ for all $\Theta \in W $.
Note that $\Theta$ is also an automorphism of subset $S_+ \oplus S_-$.
This will give us the possibility to extend Weyl reflections over the
coefficients $ g^{a_1,a_2,..a_p} $ as in the following natural way:
$$ \Theta(g^{a_1,a_2,..a_p}) \equiv g^{\theta(a_1),\theta(a_2),...\theta(a_p)}
\quad,\quad a_1,a_2,...a_p \in S_+ \oplus S_- \ \ . \eqno(II.7) $$

It could be useful to explain all the notation here with an example.
For this, let us consider, say, $A_5$ Lie algebra. As is defined above, the
generator indices take values from the sets
\vskip 3mm
\par (i)   $S_+ \equiv \{1,2, .. 15 \} $ \ \ for positive non-zero roots ,
\vskip 2mm
\par (ii)  $S_- \equiv \{ 16,17, .. 30 \}  $ \ \ for negative non-zero roots ,
\vskip 2mm
\par (iii) $S_\circ \equiv \{ 31,32, .. 35 \} $ \ \ for zero-roots .
\vskip 3mm
One has a lexicographical ordering for the composite roots in terms
of simple roots $\alpha_i$ \ (i=1,.. 5) :
$$ \eqalign{
\alpha_6 &= \alpha_1+\alpha_2 \ , \ \alpha_7 = \alpha_2+\alpha_3 \ ,
\alpha_8 = \alpha_3+\alpha_4 \ , \ \alpha_9 = \alpha_4+\alpha_5 \ , \cr
\alpha_{10} &= \alpha_1+\alpha_2 +\alpha_3 \ , \
\alpha_{11} = \alpha_2+\alpha_3 +\alpha_4 \ , \
\alpha_{12} = \alpha_3+\alpha_4 +\alpha_5 \ , \cr
\alpha_{13} &= \alpha_1+\alpha_2 +\alpha_3+\alpha_4 \ , \
\alpha_{14} = \alpha_2+\alpha_3 +\alpha_4+\alpha_5 \ , \cr
\alpha_{15} &= \alpha_1+\alpha_2 +\alpha_3+\alpha_4+\alpha_5 }  \eqno(II.8) $$
It will be instructive to show here that how invariance, parity, duality and
Weyl-symmetry properties reduce the number of coefficients before solving them
from (II.6). It is clear, for instance, that $g^{1,2,3,25}$ fulfills the
invariance property while its equivalents are
$$ \eqalign{
&g^{1,2,3,25} \sim g^{10,16,17,18} \ \ , \cr
&g^{1,2,3,25} \sim g^{3,4,5,27}  } $$
due respectively to parity and duality properties. It has also several
equivalents under the actions of Weyl-symmetry. With respect to simple
roots of $A_5$, its equivalents will be respectively
$$ \eqalign{
&g^{1,2,3,25} \sim g^{3,6,16,26} \ \ , \cr
&g^{1,2,3,25} \sim g^{6,7,17,25} \ \ , \cr
&g^{1,2,3,25} \sim g^{1,7,18,21} \ \ , \cr
&g^{1,2,3,25} \sim g^{1,2,8,28} \ \ , \cr
&g^{1,2,3,25} \sim g^{1,2,3,25}   }  $$
due to Weyl-symmetry.

All these properties restrict to some extent the number of coefficients
$ g^{A_1,A_2,..A_p} $ which are unknowns of the equations (II.6) but it is
readily seen that there are still a huge number of free parameters which
simply made the generalizations difficult. It is therefore clear that the
existence  of {\bf clusters} formed out of the coefficients $ g^{A_1,A_2,..A_p} $
{\bf having the same numerical value} is of fundamental importance here and
our main observation is that they can be determined by a set of properly
chosen indicators. In order to define these indicators for degrees p=4,5,
we need two kinds of scalar products $\kappa_1$ and $\kappa_2$. First one of
these is the usual one:
$$ \kappa_1(a,b) \equiv (\alpha_a,\alpha_b) \ \ \ \ , \ \ \ \
a,b \in S_+ \oplus S_- \ \ . \eqno(II.9) $$
For root or weight lattices of classical and exceptional Lie algebras, such
a scalar product are always defined on simple roots $\alpha_i$ by Cartan
matrix elements
$K_{ij} \equiv {2~(\alpha_i,\alpha_j) \over (\alpha_j,\alpha_j) } $ . \
This can then be extended to the whole root or weight lattice when one recalls
that roots are Z-linear and weights are Q-linear combinations of simple roots.

Our second scalar product is defined by
$$ \eqalign{
\kappa_2(i,\alpha_j + ... + \alpha_{j+k}) &= n_- \ \ , \ \ i<j \cr
\kappa_2(i,\alpha_j + ... + \alpha_{j+k}) &= n_\circ \ \ , \ \ j \leq i \leq j+k \cr
\kappa_2(i,\alpha_j + ... + \alpha_{j+k}) &= n_+ \ \ ,\ \ i>j+k \ \ . } \eqno(II.10) $$
$ n_+ , n_\circ , n_- $ here represent three different numbers.
A choice $n_-=-1~,~n_\circ=2~,~n_+=1$ will be made in following chapters.

\vskip 3mm
\par {\bf{III. FOURTH ORDER SOLUTIONS}}
\vskip 3mm
\par In chapters (III) and (IV), we assume  \
$ a_1,a_2,...\in S_+ \oplus S_- $ \ , \
$ I_1,I_2,...\in S_\circ $ and $ i_1,i_2,...\in I_\circ $. It is now useful
to study the whole set of coefficients $ g^{A_1,A_2.. A_p} $ in the following
four sub-classes:
$$ \eqalign{
&(T^{(0)}) \ \ \ \ g^{a_1,a_2,a_3,a_4}  ,  \cr
&(T^{(1)}) \ \ \ \ g^{a_1,a_2,a_3,I_1}  ,  \cr
&(T^{(2)}) \ \ \ \ g^{a_1,a_2,I_1,I_2}  ,  \cr
&(T^{(4)}) \ \ \ \ g^{I_1,I_2,I_3,I_4}  .     } \eqno(III.1)  $$
It would be helpful to recall here that indices $a_1,a_2,..$ are for non-zero
roots while $ I_1,I_2,..$ correspond to zero roots. Such a classification
could therefore be considered to be suitable because, as we emphasized above,
for the numerical values of coefficients $ g^{A_1,A_2.. A_p} $ we expect some
polinomials which are rational in N and also multi-linear in indices coming
from the zero roots only. Within such a framework, it is natural to expect one
polinomial for each one of the sub-classes in (III.1). This is trivial for
$T^{(4)}$ but the more will be seen below for the other sub-classes. It will
be seen that each sub-class $ T^{(s)} $ is a direct sum of their clusters and
each cluster is represented by a different polinomial. To specify the clusters,
following definition of indicators seem to be the most convenient ones:
$$ \eqalign{ IND(T^{(0)}) &\equiv \Sigma_0( \kappa_1(a_1,a_2),\kappa_1(a_1,a_3),\kappa_1(a_1,a_4),\kappa_1(a_2,a_3),\kappa_1(a_2,a_4),\kappa_1(a_3,a_4) ) \ \ \ \ , \cr
& \ \ \ \ \ \ \ \cr
IND(T^{(1)}) &\equiv \Sigma_1(\kappa_2(I_1,\alpha_{a_1}),\kappa_2(I_1,\alpha_{a_2}),\kappa_2(I_1,\alpha_{a_3}) ) \ \ \ \ , \cr
& \ \ \ \ \ \ \ \cr
IND(T^{(2)}) &\equiv \Sigma_2(
\Gamma(\kappa_2(I_1,\alpha_{a_1}),\kappa_2(I_1,\alpha_{a_2})) \ , \
\Gamma(\kappa_2(I_2,\alpha_{a_1}),\kappa_2(I_2,\alpha_{a_2}) ) ) \ \ .
} \eqno(III.2) $$
Let us first study the action of these indicators on $T^{(0)}$.
It is sufficient to make this in the $A_5$ example given above because
the results are independent of rank N. A set of appropriately chosen
representatives is now
$$g^{1,1,16,16} \ , \ g^{1,2,3,25} \ , \ g^{1,2,16,17} \ , \
g^{1,3,16,18} \in T^{(0)}  \eqno(III.3) $$
on which the indicators act as
$$ \eqalign{
&IND(g^{1,1,16,16}) = \Sigma_0(-2,-2,-2,-2, \ \ 2, \ \ 2) \equiv \Sigma_0(1) \cr
&IND(g^{1,2,3,25} ) \ = \Sigma_0(-1,-1,-1,-1, \ \ 0, \ \ 0) \equiv \Sigma_0(2) \cr
&IND(g^{1,2,16,17}) = \Sigma_0(-2,-2,-1,-1, \ \ 1, \ \ 1) \equiv \Sigma_0(3) \cr
&IND(g^{1,3,16,18}) = \Sigma_0(-2,-2, \ \ 0, \ \ 0, \ \ 0, \ \ 0) \equiv \Sigma_0(4) }
\eqno(III.4)  $$
The quantities $\Sigma$ are assumed to be completely symmetrical in their
indices. Calculations can be made by the aid of simple fortran-programs
for all other elements of $T^{0}$ and for any $A_N$ other than $A_5$. The
results then show us that there are nothing else other than $\Sigma_0(k)$'s
for k=1,2,3,4. We outline this fact by saying that {\bf the indicators receive
four different values on the sub-class $T^{0}$}. The following result
reflects the relevance here:

For any $A_N$, any two elements g(1),g(2) $\in T^{0} $ have the same numerical
value on condition that
$$ IND(g(1)) = IND(g(2)) \ \ .  $$
This is the main observation which reveals us the existence of sub-sets
which we would like to call {\bf clusters}.

The similar analysis shows us that indicators take 4 and 6 different values on
$T^{(1)}$ and $T^{(2)}$ respectively. To see this, it is sufficient to
consider the representatives
$$ \eqalign{
&g^{1,2,21,31} \ , \ g^{1,2,21,32} \ , \ g^{2,3,22,31} \ , \
g^{1,2,21,33} \in T^{(1)}  \cr
&g^{2,17,31,31} \ , \ g^{2,17,31,32} \ , \ g^{2,17,31,33} \ , \
g^{1,16,31,31} \ , \ g^{1,16,32,32} \ , \
g^{1,16,31,32} \in T^{(2)} } \eqno(III.5) $$
with corresponding actions
$$ \eqalign{
&IND(g^{1,2,21,31}) = \Sigma_1(\ \ 1, \ \ 2, \ \ 2) \equiv \Sigma_1( \ \ 1) \cr
&IND(g^{1,2,21,32}) = \Sigma_1(-1, \ \ 2, \ \ 2) \equiv \Sigma_1(-1) \cr
&IND(g^{2,3,22,31}) = \Sigma_1(\ \ 1, \ \ 1, \ \ 1) \equiv \Sigma_1( \ \ 2) \cr
&IND(g^{1,2,21,33}) = \Sigma_1(-1,-1,-1) \equiv \Sigma_1(-2) }
\eqno(III.6)  $$
and
$$ \eqalign{
&IND(g^{2,17,31,31}) = \Sigma_2( \ \ 1, \ \ 1) \equiv \Sigma_1( \ \ 1) \cr
&IND(g^{1,16,32,32}) = \Sigma_2(-1,-1) \equiv \Sigma_1(-1) \cr
&IND(g^{2,17,31,32}) = \Sigma_2( \ \ 1, \ \ 2) \equiv \Sigma_1( \ \ 2) \cr
&IND(g^{1,16,31,32}) = \Sigma_2(-1, \ \ 2) \equiv \Sigma_1(-2) \cr
&IND(g^{2,17,31,33}) = \Sigma_2(-1, \ \ 1) \equiv \Sigma_1( \ \ 3) \cr
&IND(g^{1,16,31,31}) = \Sigma_2( \ \ 2, \ \ 2) \equiv \Sigma_1( \ \ 4) \ \ .  }
\eqno(III.7)  $$
It is seen here that we need to define an extra generator $\Gamma$ with the
following values on sub-class $T^{(2)}$ :
$$ \eqalign{
&\Gamma( \ \ 1, \ \ 1) \equiv \Gamma( \ 1) \ , \cr
&\Gamma(-1,-1) \equiv \Gamma(-1) \ , \cr
&\Gamma( \ \ 2, \ \ 2) \equiv \Gamma( \ 2) } \eqno(III.8) $$
As a result of this discussion, for any one of the coefficients
$ g^{A_1,A_2,A_3,A_4}$ of the fourth order Casimir we have one of the
following polinomials:
$$ \eqalign{
&g^{a_1,a_2,a_3,a_4} \equiv g_4(N) ~y_k(N) \ ,
\ \ \ \ \ \ \ \ \ \ \ \ \ \ k=1,..4 \ \ \ \ , \cr
&g^{a_1,a_2,a_3,I_1} \equiv g_4(N) ~y_k(i_1,~N) \ ,
\ \ \ \ \ \ \ \ \ k=1,2 \ \ \ \ , \cr
&g^{a_1,a_2,I_1,I_2} \equiv g_4(N) ~y_k(i_1,i_2,~N) \ ,
\ \ \ \ \ \ k=1,..4 \ \ \ \ \ , \cr
&g^{I_1,I_2,I_3,I_4} \equiv g_4(N) ~y(i_1,i_2,i_3,i_4,~N) \ \ } \eqno (III.9) $$
A point which is important especially for higher order Casimirs is the
fact that the coefficients $ g^{A_1,A_2,A_3,A_4}$ are in general rational
polinomials of the rank N. It is therefore crucial to know here that
$$ g_4(N) \equiv {1 \over N ~(N-1) ~(N-2)} \ \  . \eqno(III.10)  $$
Before attacking to solve (II.6), one must also recall that the number of
coefficients with different numerical values is further reduced by the aid
of the properties mentioned in section II. To this end, it is important to
notice, in view of decompositions (III.9), that duality properties can be
given most conveniently as in the following:
$$ \eqalign{
&y_{-k}(i_1,N) = (-1)^1 ~y_k(N+1-i_1,N) \ , \ k=1,2 \ \ \ \ , \cr
\ \ \ \ \ \
&y_{-k}(i_1,i_2,N) = (-1)^2 ~y_k(N+1-i_2,N+1-i_1,N) \ , \ k=1,4  \ \ \ \ , \cr
\ \ \ \ \ \
&y(i_1,i_2,i_3,i_4,N) = (-1)^4~y(N+1-i_4,N+1-i_3,N+1-i_2,N+1-i_1,N) \ . } \eqno (III.11) $$

All these have in mind, we can easily solve equations (II.6) for only a highly
reduced number of unknown coefficients. This can be repeated for several values
of rank N which make the following generalizations possible. For this, it will
be useful to define some auxiliary polinomials $ \rho_r(N) \ , \ (r=0,.. 4) $
with the following definitions:
$$ \eqalign{
\rho_0(N)&=(N - 1)~u(1) - N~u(2) \cr
\rho_1(N)&=2~(N - 1)~(2~N - 1)~u(1) - 3~N^2~u(2) \cr
\rho_2(N)&= - (N - 1)~(N^2 - 2~N + 3)~u(1) + N~(N^2 - N + 1)~u(2) \cr
\rho_3(N)&=-2~(N - 1)~(N^2 + 2)~u(1) + N~(2~N^2 + N + 2)~u(2) \cr
\rho_4(N)&=(N - 1)~(N - 2)~((2~N - 10)~u(1) - 3~N~u(2)) \ \ \ . }
\eqno(III.12) $$
The two free parameters u(1) and u(2) here are chosen with the following
values of the unique polinomial representing elements of the subset $T^{(4)}$:
$$ y(1,1,1,1,N) \equiv u(1) \ \ , \ \ y(1,1,2,2,N) \equiv u(2) \ \ . \eqno(III.13). $$
Note here that the choices (III.13) is in correspondence with partitions
4 = 2+2. In result, the polinomials representing fourth order Casimir of
$A_N$ Lie algebras are obtained in the following forms:
$$ \eqalign{
y_1(N)&=-{1 \over 3}~\rho_4(N) \cr
y_2(N)&={1 \over 6}~(N + 1)~\rho_1(N) \cr
y_3(N)&=-{1 \over 6}~\rho_4(N) \cr
y_4(N)&=\rho_2(N) \ \ } \eqno(III.14) $$
for subset $T^{(0)}$,
$$ \eqalign{
y_1(i_1,N)&={1 \over 6}~\rho_1(N)~(N + 1 - 3~i_1) \cr
y_2(i_1,N)&=-{1 \over 2}~\rho_1(N)~i_1 \ \ } \eqno(III.15) $$
for subset $T^{(1)}$,
$$ \eqalign{
y_1(i_1,i_2,N)&=i_1~((N + 1)~i_2~\rho_0(N) + \rho_2(N)) \cr
y_2(i_1,i_2,N)&={1 \over 2}~i_1~(2~(N + 1)~i_2~\rho_0(N) + \rho_3(N)) \cr
y_3(i_1,i_2,N)&=-(N + 1)~i_1~(N + 1 - i_2)~\rho_0(N) \cr
y_4(i_1,i_2,N)&={1 \over 6}~( 6~(N + 1)~\rho_0(N)~i_1~i_2 \cr
&+ 3~\rho_3(N)~i_1 - 3~\rho_1(N)~i_2 + 2~(N + 1)~\rho_1(N)) \ \ } \eqno(III.16) $$
for subset $T^{(2)}$,
$$ y(i_1,i_2,i_3,i_4,N)=i_1~(N + 1 - i_4)~(
(3~i_2~i_3 - (N + 1)~(2~i_2 + i_3))~\rho_0(N) + \rho_1(N) )  \eqno(III.17)  $$
for subset $T^{(4)}$.

\vskip 3mm
\par {\bf{IV. FIFTH ORDER SOLUTIONS}}
\vskip 3mm
In this section  we summarize our solutions to equations (II.6) for p=5.
As being in line with the former section, we study the whole set of
coefficients $ g^{A_1,\ ...,\ A_5} $ in the following five sub-classes:
$$ \eqalign{
&(T^{(0)}) \ \ \ \ g^{a_1,a_2,a_3,a_4,a_5} \ , \cr
&(T^{(1)}) \ \ \ \ g^{a_1,a_2,a_3,a_4,I_1} \ , \cr
&(T^{(2)}) \ \ \ \ g^{a_1,a_2,a_3,I_1,I_2} \ , \cr
&(T^{(3)}) \ \ \ \ g^{a_1,a_2,I_1,I_2,I_3} \ , \cr
&(T^{(5)}) \ \ \ \ g^{I_1,I_2,I_3,I_4,I_5} \ . } \eqno(IV.1) $$
There should be no confusion between (IV.1) and (III.1) because the notations
are clear. We now want to show that these sub-classes have the following
direct sum decompositions in their clusters:
$$ \eqalign{
&T^{(0)} \equiv \bigoplus^4_{k=1} \ T^{(0)}(k) \ , \cr
&T^{(1)} \equiv \bigoplus^{13}_{k=-7} \ T^{(1)}(k) \ , \cr
&T^{(2)} \equiv \bigoplus^6_{k=-4} \ T^{(2)}(k) \ , \cr
&T^{(3)} \equiv \bigoplus^6_{k=-4} \ T^{(3)}(k) \ \ . } \eqno(IV.2)  $$
On each particular sub-class, we define the indicators act in the
following ways:
$$ \eqalign{
IND(T^{(0)}) &\equiv \Sigma_0(\kappa_1(a_1,a_2),\kappa_1(a_1,a_3),\kappa_1(a_1,a_4),\kappa_1(a_1,a_5),\kappa_1(a_2,a_3) \ , \cr
& \ \ \ \ \ \ \ \ \ \kappa_1(a_2,a_4),\kappa_1(a_2,a_5),\kappa_1(a_3,a_4),\kappa_1(a_3,a_5),\kappa_1(a_4,a_5) ) \ \ , \cr
& \ \ \ \ \ \ \ \cr
IND(T^{(1)}) &\equiv \Sigma_1(\Gamma_{11}(\kappa_1(a_1,a_2),\kappa_1(a_1,a_3),\kappa_1(a_1,a_4),\kappa_1(a_2,a_3),\kappa_1(a_2,a_4),\kappa_1(a_3,a_4)) \ , \cr
& \ \ \ \ \ \ \ \ \ \Gamma_{12}(\kappa_2(I_1,\alpha_{a_1}),\kappa_2(I_1,\alpha_{a_2}),\kappa_2(I_1,\alpha_{a_3}),\kappa_2(I_1,\alpha_{a_4}) ) ) \ \ , \cr
& \ \ \ \ \ \ \ \cr
IND(T^{(2)}) &\equiv \Sigma_2(\Gamma_2(\kappa_2(I_1,\alpha_{a_1}),\kappa_2(I_1,\alpha_{a_2}),\kappa_2(I_1,\alpha_{a_3})) \ , \cr
& \ \ \ \ \ \ \ \ \ \Gamma_2(\kappa_2(I_2,\alpha_{a_1}),\kappa_2(I_2,\alpha_{a_2}),\kappa_2(I_2,\alpha_{a_3})) ) \ \ , \cr
& \ \ \ \ \ \ \ \cr
IND(T^{(3)}) &\equiv \Sigma_3(\Gamma_3(\kappa_2(I_1,\alpha_{a_1}),\kappa_2(I_1,\alpha_{a_2})) \ , \cr
& \ \ \ \ \ \ \ \ \ \Gamma_3(\kappa_2(I_2,\alpha_{a_1}),\kappa_2(I_2,\alpha_{a_2})) \ , \cr
& \ \ \ \ \ \ \ \ \ \Gamma_3(\kappa_2(I_3,\alpha_{a_1}),\kappa_2(I_3,\alpha_{a_2})) ) \ \ . }
\eqno(IV.3)  $$
It is seen that in all these actions the scalar products are just the same as
in the 4th order. This reflects the similarity in the construction of 4th and
5th order Casimir operators of $A_N$ Lie algebras. As will be explained
elsewhere, for 6th and 7th order Casimirs we know that one needs an additional
scalar product $\kappa_3$ to define corresponding indicators.

Note also that we need to introduce several $ \Gamma$-generators in (IV.3).
Explicit calculations show that they take the following different values:
$$ \eqalign{
&\Gamma_{11}(-2,-2,-2,-2, \ \ 2, \ \ 2) \equiv \Gamma_{11}( \ 1)  \cr
&\Gamma_{11}(-1,-1,-1,-1, \ \ 0, \ \ 0) \equiv \Gamma_{11}( \ 2)  \cr
&\Gamma_{11}(-2,-2,-1,-1, \ \ 1, \ \ 1) \equiv \Gamma_{11}( \ 3)  \cr
&\Gamma_{11}(-2,-2, \ \ 0, \ \ 0, \ \ 0, \ \ 0) \equiv \Gamma_{11}( \ 4) } \eqno(IV.4) $$
$$ \eqalign{
&\Gamma_{12}( \ \ 1, \ \ 1, \ 1, \ 1) \equiv \Gamma_{12}( \ 1)    \cr
&\Gamma_{12}( \ \ 1, \ \ 1, \ 2, \ 2) \equiv \Gamma_{12}( \ 2)    \cr
&\Gamma_{12}( \ \ 2, \ \ 2, \ 2, \ 2) \equiv \Gamma_{12}( \ 3)    \cr
&\Gamma_{12}(-1, \ \ 1, \ \ 2, \ \ 2) \equiv \Gamma_{12}( \ 4)    \cr
&\Gamma_{12}(-1,-1, \ \ 1, \ \ 1) \equiv \Gamma_{12}( \ 5)        \cr
&\Gamma_{12}(-1,-1,-1,-1) \equiv \Gamma_{12}(-1)                  \cr
&\Gamma_{12}(-1,-1, \ \ 2, \ \ 2) \equiv \Gamma_{12}(-2) \ \ . } \eqno(IV.5) $$
$$ \eqalign{
&\Gamma_2( \ \ 1, \ \ 2, \ \ 2) \equiv \Gamma_2( \ 1) \ , \cr
&\Gamma_2( \ \ 1, \ \ 1, \ \ 1) \equiv \Gamma_2( \ 2) \ , \cr
&\Gamma_2(-1, \ \ 2, \ \ 2) \equiv \Gamma_2(-1) \ , \cr
&\Gamma_2(-1,-1,-1) \equiv \Gamma_2(-2)  } \eqno(IV.6)   $$
$$ \eqalign{
&\Gamma_3( \ \ 1, \ \ 1) \equiv \Gamma_3( \ 1)     \cr
&\Gamma_3( \ \ 2, \ \ 2) \equiv \Gamma_3( \ 2)     \cr
&\Gamma_3(-1,-1) \equiv \Gamma_3(-1)         } \eqno(IV.7) $$
Both $\Gamma$ and $\Sigma$-generators are assumed to be completely symmetrical
in their indices.

We can continue now in the $A_5$ example again. By using above values of
$\Gamma$-generators, one can easily show that indicators take the following
values on chosen elements of corresponding sub-classes:
$$ \eqalign{
&IND(g^{1,1,2,16,21}) = \Sigma_0(-2,-2,-1,-1,-1,-1,-1,\ \ 1,\ \ 1,\ \ 2) \equiv \Sigma_0(1) \cr
&IND(g^{1,2,3,\ 4,28}) = \Sigma_0(-1,-1,-1,-1,-1,\ \ 0,\ \ 0,\ \ 0,\ \ 0,\ \ 0) \equiv \Sigma_0(2) \cr
&IND(g^{1,2,3,16,22}) = \Sigma_0(-2,-1,-1,-1,-1,-1,\ \ 0,\ \ 0,\ \ 1,\ \ 1) \equiv \Sigma_0(3) \cr
&IND(g^{1,2,4,19,21}) = \Sigma_0(-2,-1,-1,-1,\ \ 0,\ \ 0,\ \ 0,\ \ 0,\ \ 0,\ \ 0) \equiv \Sigma_0(4) \ \ . }
\eqno(IV.8)  $$
\vskip 15mm
$$ \eqalign{
&IND(g^{1, \ 1,16,16, 32}) = \Sigma_1(1,1) \equiv \Sigma_1(\ 1)  \cr
&IND(g^{1, \ 2, \ 3,25, 34}) = \Sigma_1(2,1) \equiv \Sigma_1(\ 2)  \cr
&IND(g^{1, \ 2,16,17, 33}) = \Sigma_1(3,1) \equiv \Sigma_1(\ 3)  \cr
&IND(g^{1, \ 3,16,18, 34}) = \Sigma_1(4,1) \equiv \Sigma_1(\ 4)  \cr
&IND(g^{1, \ 2, \ 3,25, 33}) = \Sigma_1(2,2) \equiv \Sigma_1(\ 5)  \cr
&IND(g^{1, \ 2,16,17, 32}) = \Sigma_1(3,2) \equiv \Sigma_1(\ 6)  \cr
&IND(g^{1, \ 3,16,18, 33}) = \Sigma_1(4,2) \equiv \Sigma_1(\ 7)  \cr
&IND(g^{1, \ 1,16,16, 31}) = \Sigma_1(1,3) \equiv \Sigma_1(\ 8)  \cr
&IND(g^{2,10,21,22, 32}) = \Sigma_1(2,3) \equiv \Sigma_1(\ 9)  \cr
&IND(g^{1, \ 6,16,21, 31}) = \Sigma_1(3,3) \equiv \Sigma_1(10)   \cr
&IND(g^{2,10,17,25, 32}) = \Sigma_1(4,3) \equiv \Sigma_1(11)   \cr
&IND(g^{1, \ 2, \ 3,25, 32}) = \Sigma_1(2,4) \equiv \Sigma_1(12)  \cr
&IND(g^{1, \ 3,16,18, 32}) = \Sigma_1(4,5) \equiv \Sigma_1(13) \cr
& \ \ \ \ \ \ \ \ \ \ \ \cr
&IND(g^{1,2,21,32,32}) = \Sigma_2( \ \ 1,1) \equiv \Sigma_2(1)  \cr
&IND(g^{1,2,21,32,33}) = \Sigma_2( \ \ 2,1) \equiv \Sigma_2(2)  \cr
&IND(g^{1,2,21,33,33}) = \Sigma_2( \ \ 2,2) \equiv \Sigma_2(3)  \cr
&IND(g^{2,3,22,31,33}) = \Sigma_2(-2,1) \equiv \Sigma_2(4)  \cr
&IND(g^{1,2,21,31,32}) = \Sigma_2(-1,1) \equiv \Sigma_2(5)  \cr
&IND(g^{2,3,22,31,34}) = \Sigma_2(-2,2) \equiv \Sigma_2(6) \cr
& \ \ \ \ \ \ \ \ \ \ \cr
&IND(g^{1,16,32,32,32}) = \Sigma_3( \ \ 1,1,1) \equiv \Sigma_3(1)  \cr
&IND(g^{2,17,31,32,32}) = \Sigma_3(-1,2,2) \equiv \Sigma_3(2)  \cr
&IND(g^{2,17,31,33,33}) = \Sigma_3(-1,1,1) \equiv \Sigma_3(3)  \cr
&IND(g^{1,16,31,32,32}) = \Sigma_3( \ \ 1,1,2) \equiv \Sigma_3(4)  \cr
&IND(g^{1,16,31,31,31}) = \Sigma_3( \ \ 2,2,2) \equiv \Sigma_3(5)  \cr
&IND(g^{2,17,31,32,33}) = \Sigma_3(-1,1,2) \equiv \Sigma_3(6) \ \ . }
\eqno(IV.9)  $$

As in the fourth order calculations, to factor out their rational parts
is crucial to obtain polinomial expressions for coefficients
$ g^{A_1,A_2,A_3,A_4,A_5} $. This is provided by the following appropriate
assumptions:
$$ \eqalign{
g^{a_1,a_2,a_3,a_4,a_5} &\equiv g_5(N)~y_k(N) \ \ , \ \ k=1,..4 \cr
g^{a_1,a_2,a_3,a_4,I_1} &\equiv g_5(N)~y_k(i_1,N) \ \ , \ \ k=1,..13 \cr
g^{a_1,a_2,a_3,I_1,I_2} &\equiv g_5(N)~y_k(i_1,i_2,N) \ \ , \ \ k=1,..6 \cr
g^{a_1,a_2,I_1,I_2,I_3} &\equiv g_5(N)~y_k(i_1,i_2,i_3,N) \ \ , \ \ k=1,..6 \cr
g^{I_1,I_2,I_3,I_4,I_5} &\equiv g_5(N)~y(i_1,i_2,i_3,i_4,i_5,N) } \eqno(IV.10) $$
where
$$ g_5(N) \equiv {1 \over N ~(N-1) ~(N-2) ~(N-3)} \ \ . \eqno(IV.11)  $$
Note here that there should be no confusion between y-polinomials defined in
expressions (III.9) and (IV.10). In all expressions above some
$\Sigma$-generators come also together with their conjugates having negative
arguments. For instance, $\Sigma_1(1)$ has a conjugate $\Sigma_1(-1)$ whereas
there is no need to introduce a conjugate for $\Sigma_1(13)$. This in effect
is due to transformations properties of coefficients under duality
transformations mentioned in section II. These properties are properly
reflected by the following expressions:
$$ \eqalign{
&y_{-k}(i_1,N) = (-1)^1~y_k(N+1-i_1,N)
\ \ \ \ \ , \ \ \ \ \ k=1,..13 \cr
&y_{-k}(i_1,i_2,N) = (-1)^2~y_k(N+1-i_2,N+1-i_1,N)
\ \ \ , \ \ \ k=1,..6 \cr
&y_{-k}(i_1,i_2,i_3,N) = (-1)^3~y_k(N+1-i_3,N+1-i_2,N+1-i_1,N)
\ , \ k=1,..6 \cr
&y(i_1,i_2,i_3,i_4,i_5,N) =
(-1)^5~y(N+1-i_5,N+1-i_4,N+1-i_3,N+1-i_2,N+1-i_1,N) \ \ .} \eqno(IV.12) $$

As in (III.12), it would also be useful to define here some auxiliary
polinomials $ \tau_r(N) \ (r=0,.. 8) $:
$$ \eqalign{
\tau_0(N)&= - (5~N - 9)~v(1) + 5~N~v(2) \cr
\tau_1(N)&=(N - 2)~(N - 3)~((N - 7)~v(1) - 2~N~v(2)) \cr
\tau_3(N)&=(N^3 - 2~N^2 + 2~N - 3)~v(1) - N~(N^2 + 1)~v(2) \cr
\tau_2(N)&=(N + 1)~( (11~N^2 - 23~N + 6)~v(1) - 10~N^2~v(2) ) \cr
\tau_4(N)&=(N^3 + 9~N^2 - 21~N + 3)~v(1) - N~(N^2 + 10~N + 1)~v(2) \cr
\tau_5(N)&=(3~N^3 + 5~N^2 - 17~N - 3)~v(1) - N~(N + 3)~(3~N + 1)~v(2) \cr
\tau_6(N)&=-(3~N^3 + 16~N^2 - 40~N + 3) v(1) + N~(3~N^2 + 20~N + 3)~v(2) \cr
\tau_7(N)&=(9~N^3 - 7~N^2 - 5~N - 21)~v(1) - N (9~N^2 + 10~N + 9)~v(2) \cr
\tau_8(N)&=-(6~N^3 - N^2 - 11~N - 12)~v(1) + 2 N (3~N^2 + 5~N + 3)~v(2) } \eqno(IV.13) $$
for which the two free parameters v(1) and v(2) are chosen by
$$ y(1,1,1,1,1,N) \equiv v(1) \ \ ,\ \ y(1,1,1,2,2,N) \equiv v(2) \ \ .
\eqno(IV.14) $$
Note here that the choices (IV.14) are compatible with partitions \ 5 = 3+2.
In result, the polinomials assigned to sub-classes $ T^{(s)} $ will thus be
given as in the following:
$$ \eqalign{
y_1(N)&=-{1 \over 12}~(N + 1)~\tau_1(N) \cr
y_2(N)&={1 \over 24}~(N + 1)~\tau_2(N) \cr
y_3(N)&=-{1 \over 24}~(N + 1)~\tau_1(N) \cr
y_4(N)&=-{1 \over 2}~(N + 1)~\tau_3(N) \ \ } \eqno(IV.15) $$

for subset $T^{(0)}$,

\hfill\eject

$$ \eqalign{
y_1(i_1,N)&={1 \over 3}~i_1~\tau_1(N) \cr
y_2(i_1,N)&=-{1 \over 6}~i_1~\tau_2(N) \cr
y_3(i_1,N)&={1 \over 6}~i_1~\tau_1(N) \cr
y_4(i_1,N)&=2~i_1~\tau_3(N) \cr
y_5(i_1,N)&={1 \over 24}~(N + 1 - 4~i_1)~\tau_2(N) \cr
y_6(i_1,N)&=-{1 \over 24}~(N + 1 - 4~i_1)~\tau_1(N) \cr
y_7(i_1,N)&=-{1 \over 2}~(N + 1 - 4~i_1)~\tau_3(N) \cr
y_8(i_1,N)&=-{1 \over 6}~(N + 1 - 2~i_1)~\tau_1(N) \cr
y_9(i_1,N)&={1 \over 12}~(N + 1 - 2~i_1)~\tau_2(N) \cr
y_{10}(i_1,N)&=-{1 \over 12}~(N + 1 - 2~i_1)~\tau_1(N) \cr
y_{11}(i_1,N)&=-(N + 1 - 2~i_1)~\tau_3(N) \cr
y_{12}(i_1,N)&={1 \over 12}~(N + 1 - 2~i_1)~\tau_2(N) \cr
y_{13}(i_1,N)&= - (N + 1 - 2~i_1)~\tau_3(N) \ \ } \eqno(IV.16) $$

for subset $T^{(1)}$,
$$ \eqalign{
y_1(i_1,i_2,N)&={1 \over 24}~(12~i_1~i_2~\tau_4(N) + 2 \ (N + 1 - 2~i_2) \ \tau_2(N) \cr
& \ \ \ \ \ \ \ \ \ \ \ \ \ \ \ \ \ \ \ \ \ \ \ \ - 4~i_1~(N + 1) \tau_5(N) ) \cr
y_2(i_1,i_2,N)&={1 \over 12}~(6~i_1~i_2~\tau_4(N) - 2 \ i_1~(N + 1)~\tau_5(N) ) \cr
y_3(i_1,i_2,N)&={1 \over 4}~( 2~i_1~i_2~\tau_4(N) - 2 \ i_1~(N + 1)~\tau_3(N) \cr
y_4(i_1,i_2,N)&={1 \over 6}~(N + 1 - i_2)~( \tau_2(N) - 3~i_1~\tau_4(N) ) \cr
y_5(i_1,i_2,N)&={1 \over 48}~( 24~i_1~i_2~\tau_4(N) + 2 \ (3~(N + 1) - 4~i_2)~\tau_2(N) \cr
& \ \ \ \ \ \ \ \ \ \ \ \ \ \ \ \ \ \ \ \ \ \ \ \ \ \ + 8 \ i_1~(N + 1)~\tau_6(N) \ ) \cr
y_6(i_1,i_2,N)&=-{1 \over 2}~i_1~(N + 1 - i_2)~\tau_4(N) \ \ } \eqno(IV.17) $$

for subset $T^{(2)}$,

\hfill\eject

$$ \eqalign{
y_1(i_1,i_2,i_3,N)&=i_1~i_2~i_3~(N + 1)~\tau_0(N) - i_1~(N + 1 - 4~i_2 - 2~i_3)~\tau_3(N) \cr
y_2(i_1,i_2,i_3,N)&={1 \over 6}~(N + 1 - i_3) \times \cr
&(3~i_1 (N + 1 - 2~i_2)~(N + 1)~\tau_0(N) + 2~\tau_2(N) - 3~i_2~\tau_4(N) ) \cr
y_3(i_1,i_2,i_3,N)&= - (N + 1 - i_3)~( i_1~i_2~(N + 1)~\tau_0(N) + 2~i_1~\tau_3(N) ) \cr
y_4(i_1,i_2,i_3,N)&={1 \over 2}~(2~i_1~i_2~i_3~(N + 1)~\tau_0(N)  \cr
&- i_1~(3~(N + 1) - 4~i_3)~\tau_3(N) + i_1~i_2~\tau_7(N) ) \cr
y_5(i_1,i_2,i_3,N)&=-{1 \over 24}~ \times (12~i_1~i_3~(N + 1 - 2~i_2)~(N + 1)~\tau_0(N) \cr
&- \ 2 \ (3 \ (N + 1) - 4~i_3)~\tau_2(N) - 12 \ i_2~i_3~\tau_4(N) \cr
&- 4 \ i_2~(N + 1)~\tau_6(N) - 12 \ i_1~i_2~\tau_7(N) - 8 \ i_1~(N + 1)~\tau_8(N) ) \cr
y_6(i_1,i_2,i_3,N)&={1 \over 2} i_1~(N + 1 - 2~i_2)~(N + 1 - i_3)~(N + 1)~\tau_0(N) \ \ } \eqno(IV.18) $$

for subset $T^{(3)}$,
$$ \eqalign{
y(i_1,i_2,i_3,i_4,i_5,N)&=i_1~(N + 1 - i_5) \times ( \cr
&(4~i_2~i_3~i_4 - (N + 1)~( 3~i_2~i_3 + 2~i_2~i_4 + i_3~i_4 ) )~\tau_0(N) \cr
&+ \tau_2(N) - i_4~\tau_4(N) - (2~i_2 + i_3)~\tau_5(N) ) } \eqno(IV.19) $$

for subset $T^{(5)}$.

\vskip 3mm
\par {\bf{V. CONCLUSIONS}}
\vskip 3mm
In sections (III) and (IV), our general framework is outlined to
construct the {\bf most general} operatorial forms of fourth and fifth
order Casimir elements of $ A_N $ Lie algebras. The most general here means
that everything is expressed in terms of two free parameters which are
specified in (III.13) for p=4 and in (IV.14) for p=5. We have obtained some
generalizations for p=6 one interesting feature of which is that an additional
scalar product $ \kappa_3 $ is also necessary. This procedure could be
proceeded step by step beyond sixth order but we do not know at present how
many new indicators will be needed as the degree of Casimir operators
increases.

The second point which we would like to emphasize here is for the
sub-class $T^{(0)}$ of Casimir coefficients $ g^{a_1,a_2,a_3,a_4} $.
As we point out in section (III), it has the form
$$ T^{(0)} = T^{(0)}(1) \oplus T^{(0)}(2) \oplus T^{(0)}(3) \oplus
T^{(0)}(4) $$
in terms of its clusters $T^{(0)}(k)$, k=1,2,3,4. For a given value of rank \
N, $T^{(0)}$ contains totally
$$ dim(T^{(0)}) = {1 \over 2}~(N + 1)~N~(3~N^2~-~5~N~+~6) $$
number of elements while its clusters have the following dimensions:
$$ dim(T^{(0)}(1)) = 1 \ Binomial(N+1,2) \ , $$
$$ dim(T^{(0)}(2)) = 3 \ Binomial(N+1,3) \ , $$
$$ dim(T^{(0)}(3)) = 6 \ Binomial(N+1,4) \ , $$
$$ dim(T^{(0)}(4)) = 3 \ Binomial(N+1,4) \ . $$

Let us note here that these numbers are calculated in view only of
{\bf invariance} property of coefficients $ g^{a_1,a_2,a_3,a_4} $.

A similar calculation gives the following results for
$ g^{a_1,a_2,a_3,a_4,a_5} $:
$$ \eqalign{
dim(T^{(0)}(1)) &= N^3~-~(N + 1) \cr
dim(T^{(0)}(2)) &= {1 \over 5}~(N^5~-~5~N^4~+~5~N^3~+~5~N^2~-~6~N~+~5) \cr
dim(T^{(0)}(3)) &= 24 \ Binomial(N+1,4) \cr
dim(T^{(0)}(4)) &= 20 \ Binomial(N+1,5) }  $$
with
$$ dim(T^{(0)}) = {1 \over 30}~(N + 1)~N~(N - 1)~(11~N^2~-~25~N~+~36) $$
number of elements totally. For any one of these numbers, there is
a coincidence with dimensions of some Weyl orbits of $A_N$ Lie algebras.
This however could not only be a coincidence because each particular cluster
$T^{(0))}(k)$ forms an irreducible sub-space of $T^{(0))}$, that is
$$ \Theta(T^{(0))}(k)) \equiv T^{(0))}(k) \ \ , \ \  k=1,2,3,4. $$
under Weyl reflections
$$ \Theta(g^{a_1,a_2,a_3,a_4}) \equiv
g^{\theta(a_1),\theta(a_2),\theta(a_3),\theta(a_4)} \ \ ,
\ \ a_1,...a_4 \in S_+ \oplus S_-  $$
or
$$ \Theta(g^{a_1,a_2,a_3,a_4,a_5}) \equiv
g^{\theta(a_1),\theta(a_2),\theta(a_3),\theta(a_4),\theta(a_5)} \ \ ,
\ \ a_1,...a_5 \in S_+ \oplus S_-  $$
One can therefore said that indicators which are defined in expressions
(III.2) and (IV.3) are to be naturally assigned to subsets $T^{(0)}$ if
one recals here that the scalar product $\kappa_1$ is Weyl invariant.
As will be considered in another publication, this leads us to ask
some more fundamental symmetry underlying all the picture here.

As a final remark, we point out that Casimir elements of other Lie
algebras can be handled in terms of their most appropriate $A_N$ sub-algebras.
Following examples are $A_2 \ , \ A_5 \ , \ A_7 \ , \ A_8$ for respectively
$G_2 \ , \ E_6 \ , \ E_7 \ , \ E_8$. An exception is seem to be $F_4$
which is studied in terms of $B_4$ {\bf [15]} \ more conveniently.

\vskip3mm
\centerline{\bf {REFERENCES}}
\vskip3mm

\noindent [1] \ Hermann, R: Lie Groups for Physicists, W.A.Benjamin (1966) N.Y
\par Carter, R.W: Simple Groups of Lie Type, J.Wiley and sons (1972) N.Y
\vskip2mm
\noindent [2] \ Biedenharn, L.C: J.Math.Phys. 4, 436-445 (1963)
\par Perelemov A.M and Popov V.S: Sov.J.Nucl.Phys. 3, 676-680 (1966)
\par Louck J.D and Biedenharn L.C: J.Math.Phys. 11, 2368-2414 (1970)
\par Okubo S: J.Math.Phys. 16, 528-535 (1975)
\par Nwachuku C.O and Rashid M.A: J.Math.Phys. 17, 1611-1616 (1976)
\par Okubo S: J.Math.Phys. 18, 2382-2394 (1977)
\par King R.C and Qubanchi A: J.Phys.A , Math.Gen. 11, 1-7 (1978)
\par Edwards S.A: J.Math.Phys. 19, 164-167 (1978)
\par Englefield, M.J and King, R.C: J.Phys. A , Math.Gen 13, 2297-2317 (1980)
\par Berdjis F: J.Math.Phys. 22, 1851-1856 (1981)
\par Berdjis F and Beslmuller E: J.Math.Phys. 22, 1857-1860 (1981)
\par Hughes J.W.B and van der Jeugt J:J.Math.Phys. 26, 894-900 (1985)
\par Bincer A.M and Riesselman K: J.Math.Phys. 34, 5935-5941 (1993)
\vskip2mm
\noindent [3] \ Okubo S and Patera J: J.Math.Phys.24, 2722-2733 (1983)
\par Okubo S and Patera J: J.Math.Phys.25, 219-227 (1984)
\vskip2mm
\noindent [4] \ Scheunert M: J.Math.Phys. 24, 2681-2688 (1983)
\par Bincer A.M: J.Math.Phys. 24, 2546-2549 (1983)
\vskip2mm
\noindent [5] \ Gould M.D et al.: Eigenvalues of Casimir Invariants for
Type I Quantum Superalgebras,
\par q-alg/9506019
\vskip2mm
\noindent [6] \ Kac V.G: Proc.Natl.Acad.Sci. USA 81, 645-647 (1984)
\vskip2mm
\noindent [7] \ Thierry-Mieg J: Phys. Lett. B156, 199-202 (1985) and
Phys. Lett. B171, 163-169 (1986)
\vskip2mm
\noindent [8] \ Schwarz J.H: The Second String Revolution , hep-th/9607067
\vskip2mm
\noindent [9] \ Seiberg N. and Witten E. : Electric-Magnetic Duality,
Monopole Condensation and Confinement in N=2 Supersymmetric Yang-Mills Theory,
hep-th/9407087, Nucl.Phys. B426 , 19 (1994)
\par Seiberg N. and Witten E : Monopoles, Duality  and Chiral
Symmetry Breaking in N=2 Supersymmetric QCD, hep-th/9408099,Nucl. Phys. B431, 484 (1994)
\vskip2mm
\noindent [10] \ Karadayi H.R: J.Math.Phys. 25, 411-417 (1984)
\vskip2mm
\noindent [11] \ Borel A and Chevalley C: Mem.Am.Math.Soc. 14, 1 (1955)
\par Chih-Ta Yen: Sur Les Polynomes de Poincare des Groupes de Lie
Exceptionnels, Comptes Rendue Acad.Sci. Paris 628-630 (1949)
\par Chevalley C: The Betti Numbers of the Exceptional Simple Lie Groups,
Proceedings of the International Congress of Mathematicians, 2, 21-24 (1952)
\par Borel A: Ann.Math. 57, 115-207 (1953)
\par Coleman A.J: Can.J.Math 10, 349-356 (1958)
\vskip2mm
\noindent [12] \ Racah, G:Rend.Lincei 8, 108-112 (1950)
\par Racah, G:Lecture Notes reprinted in Ergebnisse Exacta Naturwissenschaften
\par 37, 28-84 (1951)
\vskip2mm
\noindent [13] \ Gruber, B and O'Raifertaigh, L: J.Math.Phy. 5, 1796-1804 (1964)
\vskip2mm
\noindent [14] \ Humhreys J.E: Introduction to Lie Algebras and Representation
Theory , Springer-Verlag (1972) N.Y.
\vskip2mm
\vskip2mm
\noindent [15] \ Bincer A.M: Casimir Operators of the Exceptional Group $F_4$:
the chain $B_4 \subset F_4 \subset D_{13} $ , {\bf hep-th-9312148}

\end